\documentclass[sn-mathphys-num]{sn-jnl}

\usepackage{geometry}
\usepackage{graphicx}%
\usepackage{multirow}%
\usepackage{amsmath,amssymb,amsfonts}%
\usepackage{amsthm}%
\usepackage{mathrsfs}%
\usepackage[title]{appendix}%
\usepackage{xcolor}%
\usepackage{textcomp}%
\usepackage{manyfoot}%
\usepackage{booktabs}%
\usepackage{algorithm}%
\usepackage{algorithmicx}%
\usepackage{algpseudocode}%
\usepackage{listings}%

\theoremstyle{thmstyleone}%
\theoremstyle{thmstyletwo}%

\theoremstyle{thmstylethree}%

\begin{document}

\title{Characterisation of polarising components at cryogenic temperature}


\author*[1]{\fnm{Thierry} \sur{Chanelière}}\email{thierry.chaneliere@neel.cnrs.fr}

\author[2]{\fnm{Alexei} D. \sur{Chepelianskii}}

\affil*[1]{Univ. Grenoble Alpes, CNRS, Grenoble INP, Institut N\'eel, 38000 Grenoble, France}

\affil[2]{LPS, Université Paris-Saclay, CNRS, UMR 8502, Orsay F-91405, France}


\abstract{

Controlling polarisation directly at low temperature is crucial  for development of optical spectroscopy techniques at sub-Kelvin temperatures, for example, in a hybrid scheme where light is fed into and collected in the cryostat by fibres that are as easy to install as electrical wiring, but where distortions in the fibre need to be compensated for by discrete polarising optical components. The latter are poorly characterised at low temperatures.
So we cool-down polarising components from room temperature to 4K and monitor the evolution of the polarisation properties in this range. We test a zero-order half-wave plate, a polarising beamsplitting cube and a dichroic polariser in the optical telecommunication range at 1.5$\mu$m. We show that the polarisation is maintained at the  $10^{-4}$ level within the whole temperature range. This is consistent with the typical thermal contraction of optical materials. This level of precision is sufficient for many optics experiments at low temperature. We argue that these experiments will allow the design of compact fibre based probes for cryogenic surfaces. 
}




\maketitle

\section{Introduction}\label{sec:intro} 

Optical techniques allow characterising the properties of a wide range of materials and quantum emitters. To cite a few, major applications include Raman spectroscopy (\cite{raman} and references therein) or microscopy (\cite{microscopy} and references therein), whose extension to low temperatures stimulated remarkable instrumental developments. Sub-Kelvin temperature optical spectroscopy can be a probe for the study of charges trapped on cryogenic surfaces, which have emerged as promising candidate for quantum mechanical applications \cite{leiderer1992electrons}. The electrons on helium system offer unrivaled mobility and several of its internal degrees of freedom are promising qubit platforms \cite{andrei2012two,monarkha2013two} : this includes surface bound states with transition energies in the 100 GHz range \cite{platzman1999quantum,collin2002microwave,konstantinov2009resonant,dykman2017ripplonic,yunusova2019coupling,kawakami2019image,chepelianskii2021many,kawakami2021relaxation,kawakami2023blueprint}, in plane confinement with energies in the GHz range \cite{rousseau2009addition,schuster2010proposal,rees2011point,bradbury2011efficient,ikegami2012evidence,rees2016structural,yang2016coupling,koolstra2019coupling,byeon2021piezoacoustics,beysengulov2024coulomb} and finally the spin degree of freedom \cite{lyon2006spin}. However, so far the influence of vibrational modes on the helium surface has not allowed to observe strong coupling of these states with external quantum circuits. It is not clear at the moment if this is a fundamental limit due to thermal ripplon vibration modes or a consequence of mechanical vibrations in the cryogenic environment \cite{beysengulov2022helium}.

It is thus important to characterise the vibrational state of the surface, and optical readout seems very promising for this application. On solid cryogenic surfaces, trapped charges on neon have recently shown reliable operation as charge qubits \cite{zavyalov2005electron,zhou2022single,zhou2024electron}. However, these surfaces contain more defects compared to the helium system and the structure of the neon surface is poorly characterised \cite{albrecht1993annealing}. Optical techniques are also very helpful for this problem; however, previous ellipsometry experiments relied on optical access cryostats \cite{sohaili2005triple}, while fibre optics would allow preforming such measurements in a dilution refrigerator in a shielded qubit environment. Finally, Kerr effect experiments can allow probing the orbital and spin magnetism of electrons on cryogenic surfaces which can represent an important step towards full control of all the quantum mechanical degrees of freedom of these systems.

In cryogenic polarisation sensitive experiments, the polarisation is generally controlled outside the cryostat. In this case, it must be assumed that the windows do not disturb the incoming or outgoing polarisation. This is never guaranteed and in any case requires a know-how for viewport installations \cite{obrien1984thermal}.

For fibre optics, the mechanical installation and thermalisation is no more complicated than the insertion of electrical cables, which avoids questions about the sealing and thermal radiation screening of the windows. The fibre polarisation behaviour during cool-down is difficult to control, but can be remedied by inserting polarising elements at low temperature to fix the polarisation in the experimental chamber and/or pre-compensate the incoming one at room temperature \cite{Mack:07, 10.1063/1.3574217, 10.1063/5.0012174}. This approach fundamentally assumes that polarising components maintain their properties at low temperature. This is not completely obvious.

Many things can happen when composite solids, as sophisticated optical elements (layered or cemented components \cite{grechushnikov1961quartz, MacNeille}), are cooled down from ambient to liquid helium temperature. The basic and most visible ones are mechanical damages, as delamination or cracks. Beyond these irreparable damages, modifications of the optical properties are also possible because of material contraction. This can be due to the contraction of the material itself modifying in turn the polarisation response of birefringent layer, as retardation plate, by converting the length change into a phase difference. Stress-induced birefringence is also possible when the active layer is sandwiched by spacers (of glass typically) making the heterogeneous assembly more sensitive to strain. Finally, stress can be amplified by the components mount, usually made of metals for good thermalisation \cite{obrien1984thermal}. Because of the possible interplay between these phenomena, we use unmounted optics in a sample-in-gas cryostat. In that sense, we test the bare component, still composed of different cemented elements as a half-wave plate (HWP) \cite{grechushnikov1961quartz, Hale:88}  or a polarising beamsplitting cube (PBS) \cite{MacNeille, Pezzaniti:94, li2000high}.

Before starting, one can discuss the contraction of optical materials from tabulated values as a point of comparison, since a change of length is readily converted in a modified optical path. This does not preclude a possible change in the refractive index, which will be discussed later. Let us start with the simple contraction effect.

The thermal contraction is defined as the normalised length difference (L$_\mathrm{293K}$ -L$_\mathrm{T})$/L$_\mathrm{293K}$ between room temperature (293K) and the temperature of interest (T) where L$_\mathrm{T}$ is the length of the material. As compared to common metals, it is low for transparent amorphous materials at 4K as silica glass $0.8\times10^{-4} $ \cite{corruccini1961thermal, ekin2006experimental} or  Pyrex \textsuperscript{TM} $5\times10^{-4}$ and generally larger for crystals as quartz $10\times10^{-4}$  at 100K \cite{corruccini1961thermal, WHITE19642} or sapphire $7.9\times10^{-4} $  at 40K \cite{arp1962thermal, ekin2006experimental}. The range for the  thermal contraction is then typically $10^{-3} $-$10^{-4}$, so we do expect a modification at the same level for the optical length. For example, the retardation of a birefringent plate, $\pi$-phase for a HWP and $2\pi \times$ [indices difference] $\times$ L$_\mathrm{T}$ $/$ [vacuum wavelength] in general, should not vary by more than $10^{-3} $-$10^{-4}$. This gives a certain confidence in controlling the polarisation at low temperature and is sufficient for most of the experiments. At the end, if the polarisation properties are solely explained by thermal contraction of the optical material, we expect a typical expansion coefficient $\displaystyle\frac{1}{L_\mathrm{T}} \frac{\partial L_\mathrm{T}}{\partial T} $ of the order of $10^{-5} $-$10^{-6}$ K$^{-1}$ between room and cryogenic temperature by simply dividing the previously discussed typical order of magnitude of the thermal contraction coefficient $10^{-3} $-$10^{-4}$ by a reference temperature range of 100K.

We left behind for a moment the possible variation of the refractive index which also influences the optical path length. The effect is not negligible, more precisely, it is often comparable to contraction with typical values of $-8\times10^{-6}$ K$^{-1}$ and $13\times10^{-6}$ K$^{-1}$ for quartz \cite{Toyoda_1983} and sapphire \cite{Malitson:62}, respectively, at room temperature. This can be considered an experimental fact and we can keep the order of magnitude of $10^{-5} $-$10^{-6}$ K$^{-1}$ as the total variation of the optical path length (thus including index and contraction). Upon reflection, this similarity can also be seen as a consequence of the Lorentz–Lorenz model \cite{Born1999}, which relates the index to the dipole density, leading to a first-order linear variation between index and length in relative values. This reasoning must be handled with caution since it systemically predicts an increase in index during contraction because of density increase leading to a negative index variation. This is clearly not the case for sapphire, without discarding the global order of magnitude.

We cool down three components, namely a zero-order half-wave plate (from CASIX Inc. \cite{casix}, see section \ref{sec:hwp}), a polarising beamsplitting cube (from CASIX Inc. \cite{casix}, see section \ref{sec:pbs}) and dichroic polariser (from CODIXX colorPol\textsuperscript{\textregistered}, see section \ref{sec:codixx}), in a liquid helium cryostat and monitor the polarisation properties as a function of temperature as described in section \ref{sec:setup}. Their design wavelength is in the fibre telecommunication range. These elements are commonly used in optics but are different by nature. Retardation plates use birefringent materials \cite{grechushnikov1961quartz}, PBS exploits the response of polarisation dependent coatings on cemented optics \cite{MacNeille} and dichroic polariser contains embedded plasmonic nanoparticules with anisotropic response \cite{colorPol}.

\section{Experimental setup}\label{sec:setup}
\begin{figure}[h]
\centering
\includegraphics[width=.8\columnwidth]{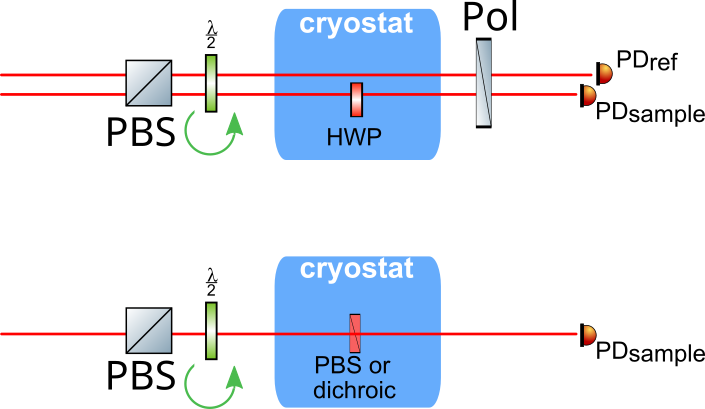} 
\caption{Top: optical configuration to record on the photodiode PD$_\mathrm{sample}$ the polarisation contrast of the HWP as a sample. We also collect the polarisation pattern of the reference beam on PD$_\mathrm{ref}$ going through the cryostat windows only. Bottom: optical configuration with PBS or the dichroic polariser as test samples. There is no need for an external polariser to directly collect the sample contrast. In both cases, we adjust the incoming polarisation by a PBS at the input and rotated by an motorised HWP.}
\label{fig:setup}
\end{figure}

For sample cooling, we operate a Janis Research SVT-200 with an optical tail. The transmission path goes through of three pairs of quartz windows (outer jacket, radiation shield and cold chamber). The optical components under test are inserted in the sample space and cooled down by a moderate contact with the sample holder and by the helium gas flowing through the variable-temperature-insert.
The laser is separated in two beams both polarised before passing through the cryostat (Fig.\ref{fig:setup}). One goes through the sample under test and the other one serves as a reference to evaluate the windows polarisation response. We rotate the incoming polarisation with a motorised mount and essentially record the polarisation contrast between the bright and extinction angles when the outgoing analyser is, respectively, parallel or crossed with respect to the incoming polarisation. Only one beam goes through the sample and the other one is used as a reference to distinguish the change due the cryostat windows. There are two possible configurations. First, with the HWP under test, an extra fixed polariser is positioned after the cryostat to acquire the polarisation contrast  (Fig.\ref{fig:setup}, top). Second, with the PBS or the dichroic polariser, the test sample in the cryostat being a polariser, the fringe pattern is directly recorded at the output without external polariser (Fig.\ref{fig:setup}, bottom). During the cool down, for a given temperature, we rotate the incoming polarisation and measure the contrast. The evolution of the contrast as a function of the temperature reveals the change of the polarising components under test from room temperature to typically 4K.

The laser wavelength (1536\,nm) falls in the fibre telecommunication range where lasers and optical components are widely available at a moderate price. Since we measure relative changes of the optical path length that modifies the polarisation, our results also apply to other optical wavelength when comparable components are available.

\section{Zero-order half-wave plate}\label{sec:hwp}

\begin{figure}[h]
\centering
\includegraphics[width=\columnwidth]{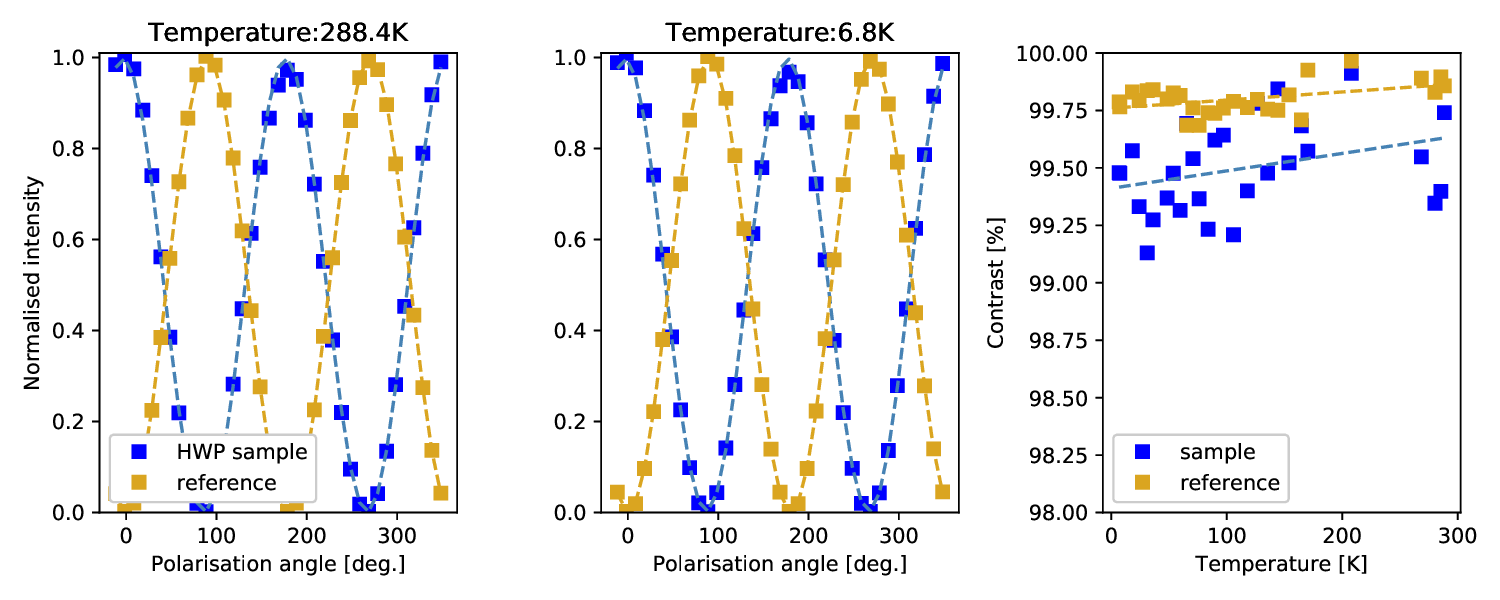} 
\caption{Left, middle: Half-wave plate polarisation contrast for a rotating incoming polarisation. The sample under test is compared to a reference (see text for details). Left and middle correspond to temperatures of 288.4K and 6.8K respectively. No major change is observed between room and cryogenic temperature. Right: Contrast of the half-wave plate and the reference beam as a function of the temperature.}
\label{fig:fringes_hwp}
\end{figure}

We first test a zero-order half-wave plate from room temperature to 6K and record the fringes contrast as the incoming polarisation is rotated (Fig.\ref{fig:fringes_hwp}).

We additionally acquire the same pattern for a reference beam passing though the cryostat windows only (Fig.\ref{fig:setup}, top). The fringe contrast is larger 99\% which is sufficient for our analysis in temperature. The extinction is here limited anyway to 99.9\% by the incoming polarisation emerging from the PBS. It degrades further because of the residual birefringence of the cryostat windows. As we will see, this is not a limiting factor for our measurement because we observe $10^{-3} $ shot-to-shot fluctuations of the contrast during the cool down and acquisition which takes a few hours.


To quantify the dependency  in the whole temperature range, we plot the evolution of the contrast in Fig.\ref{fig:fringes_hwp} (right). The sample exhibits a relatively weak dependency with an average slope $(7\pm4)\times10^{-6} $ K$^{-1}$ (linear regression as a dashed line in the figure) that should be compared to the reference average slope of $(3\pm1)\times10^{-6} $ K$^{-1}$. This latter may be explained by a residual birefringence of the cryostat windows that varies with the temperature. Concerning the sample, despite a relatively large error bar, few $10^{-6} $ K$^{-1}$ appears a realistic order of magnitude for the thermal variation of the HWP polarisation. Here we broadly confirm our preliminary analysis made in the introduction \ref{sec:intro}. We will discuss it again in conclusion \ref{sec:ccl}.

\section{Polarising beamsplitting cube}\label{sec:pbs}
\begin{figure}[h]
\centering
\includegraphics[width=\columnwidth]{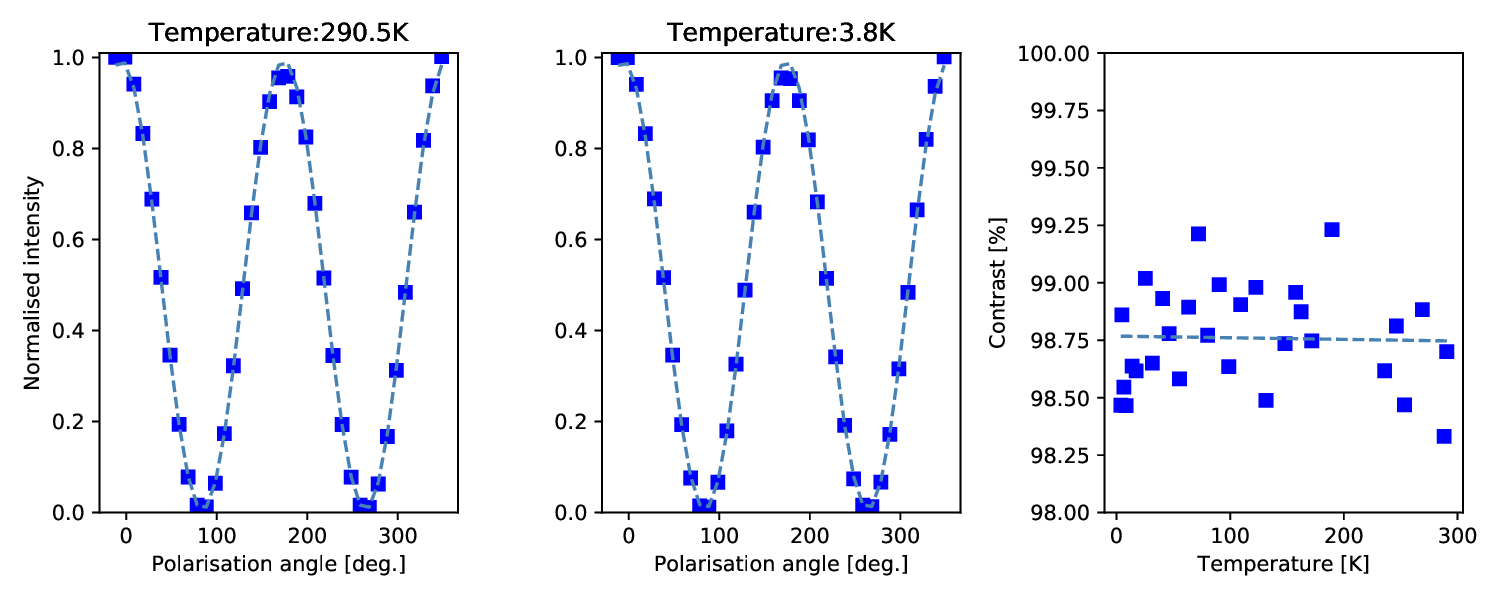} 
\caption{Left, middle: polarising beamsplitting cube contrast for a rotating incoming polarisation for the temperatures of 290.5K and 3.8K respectively. Right: Contrast as a function of the temperature.}
\label{fig:fringes_pbs}
\end{figure}

We reproduce the previous measurement with a polarising beamsplitting cube as a second test sample. There is no need to insert a polariser after the cryostat because the PBS under test serves as an analyser (Fig.\ref{fig:setup}, bottom). We simply record the outgoing intensity contrast when we rotate the incoming polarisation in Fig.\ref{fig:fringes_pbs}.


There is no observable temperature dependency over the complete range. The average slope $(-0.7\pm4)\times10^{-6} $ K$^{-1}$ is essentially zero within the error bars (Fig.\ref{fig:fringes_pbs}, right). If there is any variation, it should be lower than few $10^{-6} $ K$^{-1}$. This is consistent with the previous measurement in section \ref{sec:hwp}, both elements being cemented optics in which the polarisation active layer is sandwiched between glass spacers.


\section{Dichroic polariser}\label{sec:codixx}
\begin{figure}[h]
\centering
\includegraphics[width=\columnwidth]{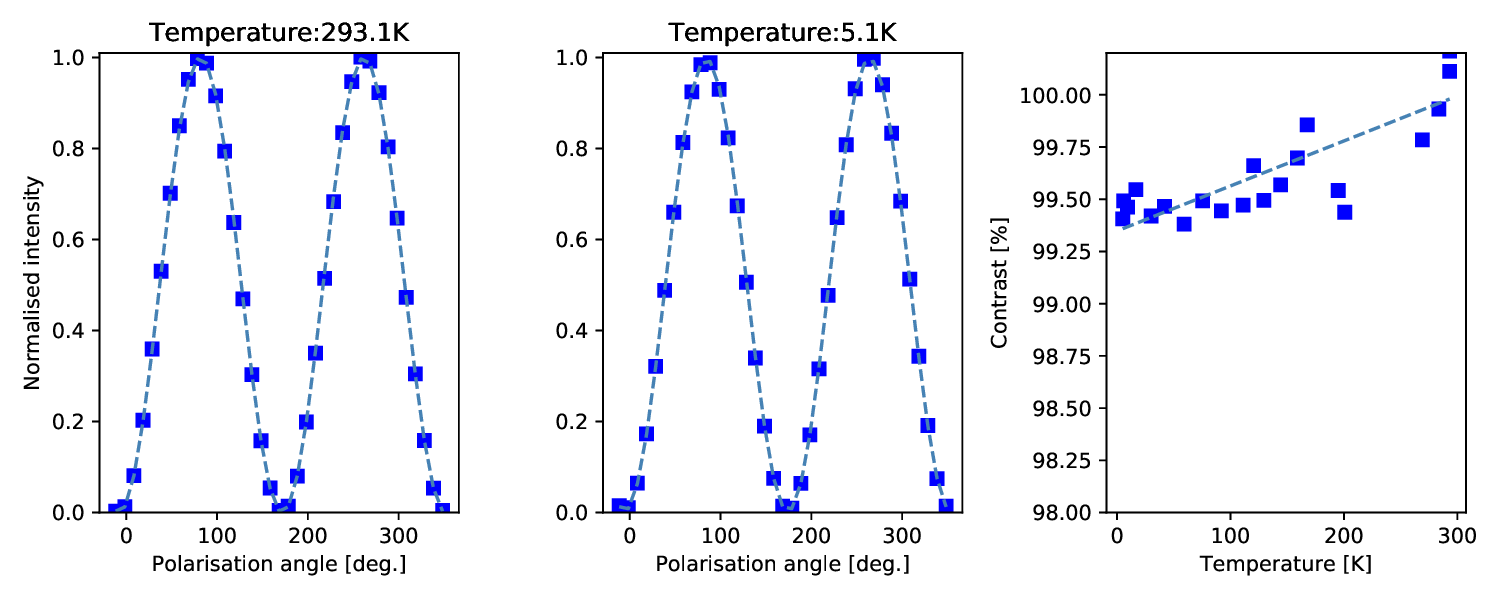} 
\caption{Left, middle: Dichroic polariser contrast for a rotating incoming polarisation for the temperatures of 293.1K and 5.1K respectively. Right: Contrast as a function of the temperature.}
\label{fig:contrast_codixx}
\end{figure}

We conclude our analysis with a dichroic polariser in which anisotropic plasmonic nanoparticules are embedded \cite{colorPol}. The technology involved is then structurally different than the previous studied HWP and PBS. As in section \ref{sec:pbs}, we record the outgoing intensity contrast in Fig.\ref{fig:contrast_codixx}.

The component design allows to obtain very large contrast $>$ 99.9\%, our typical measurement limit. We nevertheless observe a significant variation over the temperature range $(21\pm3)\times10^{-6} $ K$^{-1}$. This is larger than the other components, even if a good extinction is still preserved $>$ 99.3\%, which can be sufficient for many experiments.

The relatively large measured variation, in the $10^{-5} $ K$^{-1}$ range, should be considered with precaution. The sample has a relatively small size $5\times5$ mm that can be compared to the beam size ($\sim$ 250$\mu$m waist) and more importantly to the cryostat sample rod contraction during cool down ($\sim$ 2mm). The beam can simply move away and by-pass the polarising sample, thus exhibiting an apparent reduction of the contrast because of this misalignment. We periodically readjust the cryostat sample rod length to compensate for this, but because of the dichroic polariser reduced size and the large polarisation contrast, this sample is particularly sensitive to misalignment. The variation $(21\pm3)\times10^{-6} $ K$^{-1}$ should then be taken as an upper-bound.

\section{Conclusion}\label{sec:ccl}
Our results show that the three tested components exhibit comparable polarisation properties in overall. They maintain a good precision of the polarisation at low temperatures.

The first two samples, zero-order half-wave plate and polarising beamsplitting cube, belong to the same family of cemented optics. They do not break during cool down and they show a polarisation contrast variation of few $10^{-6} $ K$^{-1}$. This is consistent with the typical thermal contraction of optical materials like glass or optical crystals. 

The third sample, a dichroic polariser, is fundamentally different from the other two. We observe a thermal variation that is an order of magnitude larger, which may be explained by a larger sensitivity to misalignment because of the small sample size and high polarisation contrast. Despite the larger thermal variation, the polarisation contrast remains high over the whole temperature range.

Although many things can happen during cool down, such as delamination, cracks, and strain, we never observed any major degradation or ageing of components during the few complete cooling cycles from room temperature in our study.
The conclusion of our analysis during the cool down is the observation of a weak and trackable modification of the polarisation properties at low temperature.
To reduce strain that may be induced by an optics mount, we intentionally used unmounted optics that are cooled by helium gas in the variable temperature insert. This allowed us to extract the intrinsic variation of the test samples. This aspect is important to consider when the cold test chamber is under vacuum (in cryocoolers for example) and when the optical elements are thermalised by metallic intermediate mounts.

The development of specific components for low temperature seems to leave little margins for improvement. The modification of the polarisation properties depends in the first approximation on the contraction coefficient. It is therefore not possible to play on the size of the active material to reduce this dependency. On the other hand, the thinness of the current polarising layers, for example for a zero-order as opposed to a multi-order plate, is probably an advantage in ensuring the mechanical resistance of the assembly during cooling cycles. The use of low-expansion optical materials also appears to be a dead end, as these materials are usually a mixture of amorphous material and thus lose their crystalline optical properties. The introductory discussion, however, highlighted the potentially antagonistic effect of contraction (length reduction) and index increase for the material qualitatively following the Lorentz–Lorenz law as quartz. One could imagine the two compensating each other to show a reduced sensitivity of the optical path (as the product of the index by length). This definitely requires further study.

\backmatter

\bmhead{Acknowledgements}

The authors acknowledge support from the French National Research Agency (ANR) project MARS No. ANR-20-CE92-0041 and from the Plan France 2030 project QMEMO No. ANR-22-PETQ-0010.

We thank Ivan Breslavetz for the discussions and encouragements.

\bibliography{polar_cryo}


\begin{thebibliography}{49}
\ifx \bisbn   \undefined \def \bisbn  #1{ISBN #1}\fi
\ifx \binits  \undefined \def \binits#1{#1}\fi
\ifx \bauthor  \undefined \def \bauthor#1{#1}\fi
\ifx \batitle  \undefined \def \batitle#1{#1}\fi
\ifx \bjtitle  \undefined \def \bjtitle#1{#1}\fi
\ifx \bvolume  \undefined \def \bvolume#1{\textbf{#1}}\fi
\ifx \byear  \undefined \def \byear#1{#1}\fi
\ifx \bissue  \undefined \def \bissue#1{#1}\fi
\ifx \bfpage  \undefined \def \bfpage#1{#1}\fi
\ifx \blpage  \undefined \def \blpage #1{#1}\fi
\ifx \burl  \undefined \def \burl#1{\textsf{#1}}\fi
\ifx \doiurl  \undefined \def \doiurl#1{\url{https://doi.org/#1}}\fi
\ifx \betal  \undefined \def \betal{\textit{et al.}}\fi
\ifx \binstitute  \undefined \def \binstitute#1{#1}\fi
\ifx \binstitutionaled  \undefined \def \binstitutionaled#1{#1}\fi
\ifx \bctitle  \undefined \def \bctitle#1{#1}\fi
\ifx \beditor  \undefined \def \beditor#1{#1}\fi
\ifx \bpublisher  \undefined \def \bpublisher#1{#1}\fi
\ifx \bbtitle  \undefined \def \bbtitle#1{#1}\fi
\ifx \bedition  \undefined \def \bedition#1{#1}\fi
\ifx \bseriesno  \undefined \def \bseriesno#1{#1}\fi
\ifx \blocation  \undefined \def \blocation#1{#1}\fi
\ifx \bsertitle  \undefined \def \bsertitle#1{#1}\fi
\ifx \bsnm \undefined \def \bsnm#1{#1}\fi
\ifx \bsuffix \undefined \def \bsuffix#1{#1}\fi
\ifx \bparticle \undefined \def \bparticle#1{#1}\fi
\ifx \barticle \undefined \def \barticle#1{#1}\fi
\bibcommenthead
\ifx \bconfdate \undefined \def \bconfdate #1{#1}\fi
\ifx \botherref \undefined \def \botherref #1{#1}\fi
\ifx \url \undefined \def \url#1{\textsf{#1}}\fi
\ifx \bchapter \undefined \def \bchapter#1{#1}\fi
\ifx \bbook \undefined \def \bbook#1{#1}\fi
\ifx \bcomment \undefined \def \bcomment#1{#1}\fi
\ifx \oauthor \undefined \def \oauthor#1{#1}\fi
\ifx \citeauthoryear \undefined \def \citeauthoryear#1{#1}\fi
\ifx \endbibitem  \undefined \def \endbibitem {}\fi
\ifx \bconflocation  \undefined \def \bconflocation#1{#1}\fi
\ifx \arxivurl  \undefined \def \arxivurl#1{\textsf{#1}}\fi
\csname PreBibitemsHook\endcsname

\bibitem[\protect\citeauthoryear{Tian et~al.}{2016}]{raman}
\begin{barticle}
\bauthor{\bsnm{Tian}, \binits{Y.}},
\bauthor{\bsnm{Reijnders}, \binits{A.A.}},
\bauthor{\bsnm{Osterhoudt}, \binits{G.B.}},
\bauthor{\bsnm{Valmianski}, \binits{I.}},
\bauthor{\bsnm{Ramirez}, \binits{J.G.}},
\bauthor{\bsnm{Urban}, \binits{C.}},
\bauthor{\bsnm{Zhong}, \binits{R.}},
\bauthor{\bsnm{Schneeloch}, \binits{J.}},
\bauthor{\bsnm{Gu}, \binits{G.}},
\bauthor{\bsnm{Henslee}, \binits{I.}},
\bauthor{\bsnm{Burch}, \binits{K.S.}}:
\batitle{{Low vibration high numerical aperture automated variable temperature
  Raman microscope}}.
\bjtitle{Review of Scientific Instruments}
\bvolume{87}(\bissue{4}),
\bfpage{043105}
(\byear{2016})
\doiurl{10.1063/1.4944559}
{\href{https://arxiv.org/abs/https://pubs.aip.org/aip/rsi/article-pdf/doi/10.1063/1.4944559/14721225/043105\_1\_online.pdf}{{https://pubs.aip.org/aip/rsi/article-pdf/doi/10.1063/1.4944559/14721225/043105\_1\_online.pdf}}}
\end{barticle}
\endbibitem

\bibitem[\protect\citeauthoryear{Högele et~al.}{2008}]{microscopy}
\begin{barticle}
\bauthor{\bsnm{Högele}, \binits{A.}},
\bauthor{\bsnm{Seidl}, \binits{S.}},
\bauthor{\bsnm{Kroner}, \binits{M.}},
\bauthor{\bsnm{Karrai}, \binits{K.}},
\bauthor{\bsnm{Schulhauser}, \binits{C.}},
\bauthor{\bsnm{Sqalli}, \binits{O.}},
\bauthor{\bsnm{Scrimgeour}, \binits{J.}},
\bauthor{\bsnm{Warburton}, \binits{R.J.}}:
\batitle{{Fiber-based confocal microscope for cryogenic spectroscopy}}.
\bjtitle{Review of Scientific Instruments}
\bvolume{79}(\bissue{2}),
\bfpage{023709}
(\byear{2008})
\doiurl{10.1063/1.2885681}
{\href{https://arxiv.org/abs/https://pubs.aip.org/aip/rsi/article-pdf/doi/10.1063/1.2885681/14115426/023709\_1\_online.pdf}{{https://pubs.aip.org/aip/rsi/article-pdf/doi/10.1063/1.2885681/14115426/023709\_1\_online.pdf}}}
\end{barticle}
\endbibitem

\bibitem[\protect\citeauthoryear{Leiderer}{1992}]{leiderer1992electrons}
\begin{barticle}
\bauthor{\bsnm{Leiderer}, \binits{P.}}:
\batitle{Electrons at the surface of quantum systems}.
\bjtitle{Journal of Low Temperature Physics}
\bvolume{87},
\bfpage{247}--\blpage{278}
(\byear{1992})
\end{barticle}
\endbibitem

\bibitem[\protect\citeauthoryear{Andrei}{1997}]{andrei2012two}
\begin{bbook}
\bauthor{\bsnm{Andrei}, \binits{E.Y.}}:
\bbtitle{Two-Dimensional Electron Systems: on Helium and Other Cryogenic
  Substrates}.
\bpublisher{Springer},
\blocation{Netherlands}
(\byear{1997}).
\doiurl{10.1007/978-94-015-1286-2} .
\burl{http://dx.doi.org/10.1007/978-94-015-1286-2}
\end{bbook}
\endbibitem

\bibitem[\protect\citeauthoryear{Monarkha and Kono}{2004}]{monarkha2013two}
\begin{bbook}
\bauthor{\bsnm{Monarkha}, \binits{Y.}},
\bauthor{\bsnm{Kono}, \binits{K.}}:
\bbtitle{Two-Dimensional Coulomb Liquids and Solids}.
\bpublisher{Springer},
\blocation{Berlin Heidelberg}
(\byear{2004}).
\doiurl{10.1007/978-3-662-10639-6} .
\burl{http://dx.doi.org/10.1007/978-3-662-10639-6}
\end{bbook}
\endbibitem

\bibitem[\protect\citeauthoryear{Platzman and
  Dykman}{1999}]{platzman1999quantum}
\begin{barticle}
\bauthor{\bsnm{Platzman}, \binits{P.}},
\bauthor{\bsnm{Dykman}, \binits{M.}}:
\batitle{Quantum computing with electrons floating on liquid helium}.
\bjtitle{Science}
\bvolume{284}(\bissue{5422}),
\bfpage{1967}--\blpage{1969}
(\byear{1999})
\end{barticle}
\endbibitem

\bibitem[\protect\citeauthoryear{Collin et~al.}{2002}]{collin2002microwave}
\begin{barticle}
\bauthor{\bsnm{Collin}, \binits{E.}},
\bauthor{\bsnm{Bailey}, \binits{W.}},
\bauthor{\bsnm{Fozooni}, \binits{P.}},
\bauthor{\bsnm{Frayne}, \binits{P.}},
\bauthor{\bsnm{Glasson}, \binits{P.}},
\bauthor{\bsnm{Harrabi}, \binits{K.}},
\bauthor{\bsnm{Lea}, \binits{M.}},
\bauthor{\bsnm{Papageorgiou}, \binits{G.}}:
\batitle{Microwave saturation of the rydberg states of electrons on helium}.
\bjtitle{Physical review letters}
\bvolume{89}(\bissue{24}),
\bfpage{245301}
(\byear{2002})
\end{barticle}
\endbibitem

\bibitem[\protect\citeauthoryear{Konstantinov
  et~al.}{2009}]{konstantinov2009resonant}
\begin{barticle}
\bauthor{\bsnm{Konstantinov}, \binits{D.}},
\bauthor{\bsnm{Dykman}, \binits{M.}},
\bauthor{\bsnm{Lea}, \binits{M.}},
\bauthor{\bsnm{Monarkha}, \binits{Y.}},
\bauthor{\bsnm{Kono}, \binits{K.}}:
\batitle{Resonant correlation-induced optical bistability in an electron system
  on liquid helium}.
\bjtitle{Physical review letters}
\bvolume{103}(\bissue{9}),
\bfpage{096801}
(\byear{2009})
\end{barticle}
\endbibitem

\bibitem[\protect\citeauthoryear{Dykman et~al.}{2017}]{dykman2017ripplonic}
\begin{barticle}
\bauthor{\bsnm{Dykman}, \binits{M.}},
\bauthor{\bsnm{Kono}, \binits{K.}},
\bauthor{\bsnm{Konstantinov}, \binits{D.}},
\bauthor{\bsnm{Lea}, \binits{M.}}:
\batitle{Ripplonic lamb shift for electrons on liquid helium}.
\bjtitle{Physical review letters}
\bvolume{119}(\bissue{25}),
\bfpage{256802}
(\byear{2017})
\end{barticle}
\endbibitem

\bibitem[\protect\citeauthoryear{Yunusova et~al.}{2019}]{yunusova2019coupling}
\begin{barticle}
\bauthor{\bsnm{Yunusova}, \binits{K.M.}},
\bauthor{\bsnm{Konstantinov}, \binits{D.}},
\bauthor{\bsnm{Bouchiat}, \binits{H.}},
\bauthor{\bsnm{Chepelianskii}, \binits{A.}}:
\batitle{Coupling between rydberg states and landau levels of electrons trapped
  on liquid helium}.
\bjtitle{Physical Review Letters}
\bvolume{122}(\bissue{17}),
\bfpage{176802}
(\byear{2019})
\end{barticle}
\endbibitem

\bibitem[\protect\citeauthoryear{Kawakami et~al.}{2019}]{kawakami2019image}
\begin{barticle}
\bauthor{\bsnm{Kawakami}, \binits{E.}},
\bauthor{\bsnm{Elarabi}, \binits{A.}},
\bauthor{\bsnm{Konstantinov}, \binits{D.}}:
\batitle{Image-charge detection of the rydberg states of surface electrons on
  liquid helium}.
\bjtitle{Physical Review Letters}
\bvolume{123}(\bissue{8}),
\bfpage{086801}
(\byear{2019})
\end{barticle}
\endbibitem

\bibitem[\protect\citeauthoryear{Chepelianskii
  et~al.}{2021}]{chepelianskii2021many}
\begin{barticle}
\bauthor{\bsnm{Chepelianskii}, \binits{A.}},
\bauthor{\bsnm{Konstantinov}, \binits{D.}},
\bauthor{\bsnm{Dykman}, \binits{M.}}:
\batitle{Many-electron system on helium and color center spectroscopy}.
\bjtitle{Physical Review Letters}
\bvolume{127}(\bissue{1}),
\bfpage{016801}
(\byear{2021})
\end{barticle}
\endbibitem

\bibitem[\protect\citeauthoryear{Kawakami
  et~al.}{2021}]{kawakami2021relaxation}
\begin{barticle}
\bauthor{\bsnm{Kawakami}, \binits{E.}},
\bauthor{\bsnm{Elarabi}, \binits{A.}},
\bauthor{\bsnm{Konstantinov}, \binits{D.}}:
\batitle{Relaxation of the excited rydberg states of surface electrons on
  liquid helium}.
\bjtitle{Physical Review Letters}
\bvolume{126}(\bissue{10}),
\bfpage{106802}
(\byear{2021})
\end{barticle}
\endbibitem

\bibitem[\protect\citeauthoryear{Kawakami et~al.}{2023}]{kawakami2023blueprint}
\begin{barticle}
\bauthor{\bsnm{Kawakami}, \binits{E.}},
\bauthor{\bsnm{Chen}, \binits{J.}},
\bauthor{\bsnm{Benito}, \binits{M.}},
\bauthor{\bsnm{Konstantinov}, \binits{D.}}:
\batitle{Blueprint for quantum computing using electrons on helium}.
\bjtitle{Physical Review Applied}
\bvolume{20}(\bissue{5}),
\bfpage{054022}
(\byear{2023})
\end{barticle}
\endbibitem

\bibitem[\protect\citeauthoryear{Rousseau et~al.}{2009}]{rousseau2009addition}
\begin{barticle}
\bauthor{\bsnm{Rousseau}, \binits{E.}},
\bauthor{\bsnm{Ponarin}, \binits{D.}},
\bauthor{\bsnm{Hristakos}, \binits{L.}},
\bauthor{\bsnm{Avenel}, \binits{O.}},
\bauthor{\bsnm{Varoquaux}, \binits{E.}},
\bauthor{\bsnm{Mukharsky}, \binits{Y.}}:
\batitle{Addition spectra of wigner islands of electrons on superfluid helium}.
\bjtitle{Physical Review B—Condensed Matter and Materials Physics}
\bvolume{79}(\bissue{4}),
\bfpage{045406}
(\byear{2009})
\end{barticle}
\endbibitem

\bibitem[\protect\citeauthoryear{Schuster et~al.}{2010}]{schuster2010proposal}
\begin{barticle}
\bauthor{\bsnm{Schuster}, \binits{D.}},
\bauthor{\bsnm{Fragner}, \binits{A.}},
\bauthor{\bsnm{Dykman}, \binits{M.}},
\bauthor{\bsnm{Lyon}, \binits{S.}},
\bauthor{\bsnm{Schoelkopf}, \binits{R.}}:
\batitle{Proposal for manipulating and detecting spin and orbital states of
  trapped electrons on helium using cavity quantum electrodynamics}.
\bjtitle{Physical review letters}
\bvolume{105}(\bissue{4}),
\bfpage{040503}
(\byear{2010})
\end{barticle}
\endbibitem

\bibitem[\protect\citeauthoryear{Rees et~al.}{2011}]{rees2011point}
\begin{barticle}
\bauthor{\bsnm{Rees}, \binits{D.G.}},
\bauthor{\bsnm{Kuroda}, \binits{I.}},
\bauthor{\bsnm{Marrache-Kikuchi}, \binits{C.A.}},
\bauthor{\bsnm{H{\"o}fer}, \binits{M.}},
\bauthor{\bsnm{Leiderer}, \binits{P.}},
\bauthor{\bsnm{Kono}, \binits{K.}}:
\batitle{Point-contact transport properties of strongly correlated electrons on
  liquid helium}.
\bjtitle{Physical review letters}
\bvolume{106}(\bissue{2}),
\bfpage{026803}
(\byear{2011})
\end{barticle}
\endbibitem

\bibitem[\protect\citeauthoryear{Bradbury et~al.}{2011}]{bradbury2011efficient}
\begin{barticle}
\bauthor{\bsnm{Bradbury}, \binits{F.}},
\bauthor{\bsnm{Takita}, \binits{M.}},
\bauthor{\bsnm{Gurrieri}, \binits{T.}},
\bauthor{\bsnm{Wilkel}, \binits{K.}},
\bauthor{\bsnm{Eng}, \binits{K.}},
\bauthor{\bsnm{Carroll}, \binits{M.}},
\bauthor{\bsnm{Lyon}, \binits{S.A.}}:
\batitle{Efficient clocked electron transfer on superfluid helium}.
\bjtitle{Physical review letters}
\bvolume{107}(\bissue{26}),
\bfpage{266803}
(\byear{2011})
\end{barticle}
\endbibitem

\bibitem[\protect\citeauthoryear{Ikegami et~al.}{2012}]{ikegami2012evidence}
\begin{barticle}
\bauthor{\bsnm{Ikegami}, \binits{H.}},
\bauthor{\bsnm{Akimoto}, \binits{H.}},
\bauthor{\bsnm{Rees}, \binits{D.G.}},
\bauthor{\bsnm{Kono}, \binits{K.}}:
\batitle{Evidence for reentrant melting in a quasi-one-dimensional wigner
  crystal}.
\bjtitle{Physical Review Letters}
\bvolume{109}(\bissue{23}),
\bfpage{236802}
(\byear{2012})
\end{barticle}
\endbibitem

\bibitem[\protect\citeauthoryear{Rees et~al.}{2016}]{rees2016structural}
\begin{barticle}
\bauthor{\bsnm{Rees}, \binits{D.G.}},
\bauthor{\bsnm{Beysengulov}, \binits{N.R.}},
\bauthor{\bsnm{Teranishi}, \binits{Y.}},
\bauthor{\bsnm{Tsao}, \binits{C.-S.}},
\bauthor{\bsnm{Yeh}, \binits{S.-S.}},
\bauthor{\bsnm{Chiu}, \binits{S.-P.}},
\bauthor{\bsnm{Lin}, \binits{Y.-H.}},
\bauthor{\bsnm{Tayurskii}, \binits{D.A.}},
\bauthor{\bsnm{Lin}, \binits{J.-J.}},
\bauthor{\bsnm{Kono}, \binits{K.}}:
\batitle{Structural order and melting of a quasi-one-dimensional electron
  system}.
\bjtitle{Physical Review B}
\bvolume{94}(\bissue{4}),
\bfpage{045139}
(\byear{2016})
\end{barticle}
\endbibitem

\bibitem[\protect\citeauthoryear{Yang et~al.}{2016}]{yang2016coupling}
\begin{barticle}
\bauthor{\bsnm{Yang}, \binits{G.}},
\bauthor{\bsnm{Fragner}, \binits{A.}},
\bauthor{\bsnm{Koolstra}, \binits{G.}},
\bauthor{\bsnm{Ocola}, \binits{L.}},
\bauthor{\bsnm{Czaplewski}, \binits{D.}},
\bauthor{\bsnm{Schoelkopf}, \binits{R.}},
\bauthor{\bsnm{Schuster}, \binits{D.}}:
\batitle{Coupling an ensemble of electrons on superfluid helium to a
  superconducting circuit}.
\bjtitle{Physical Review X}
\bvolume{6}(\bissue{1}),
\bfpage{011031}
(\byear{2016})
\end{barticle}
\endbibitem

\bibitem[\protect\citeauthoryear{Koolstra et~al.}{2019}]{koolstra2019coupling}
\begin{barticle}
\bauthor{\bsnm{Koolstra}, \binits{G.}},
\bauthor{\bsnm{Yang}, \binits{G.}},
\bauthor{\bsnm{Schuster}, \binits{D.I.}}:
\batitle{Coupling a single electron on superfluid helium to a superconducting
  resonator}.
\bjtitle{Nature communications}
\bvolume{10}(\bissue{1}),
\bfpage{5323}
(\byear{2019})
\end{barticle}
\endbibitem

\bibitem[\protect\citeauthoryear{Byeon et~al.}{2021}]{byeon2021piezoacoustics}
\begin{barticle}
\bauthor{\bsnm{Byeon}, \binits{H.}},
\bauthor{\bsnm{Nasyedkin}, \binits{K.}},
\bauthor{\bsnm{Lane}, \binits{J.}},
\bauthor{\bsnm{Beysengulov}, \binits{N.}},
\bauthor{\bsnm{Zhang}, \binits{L.}},
\bauthor{\bsnm{Loloee}, \binits{R.}},
\bauthor{\bsnm{Pollanen}, \binits{J.}}:
\batitle{Piezoacoustics for precision control of electrons floating on helium}.
\bjtitle{Nature Communications}
\bvolume{12}(\bissue{1}),
\bfpage{4150}
(\byear{2021})
\end{barticle}
\endbibitem

\bibitem[\protect\citeauthoryear{Beysengulov
  et~al.}{2024}]{beysengulov2024coulomb}
\begin{barticle}
\bauthor{\bsnm{Beysengulov}, \binits{N.R.}},
\bauthor{\bsnm{Sch{\o}yen}, \binits{{\O}.S.}},
\bauthor{\bsnm{Bilek}, \binits{S.D.}},
\bauthor{\bsnm{Flaten}, \binits{J.B.}},
\bauthor{\bsnm{Leinonen}, \binits{O.}},
\bauthor{\bsnm{Hjorth-Jensen}, \binits{M.}},
\bauthor{\bsnm{Pollanen}, \binits{J.}},
\bauthor{\bsnm{Kristiansen}, \binits{H.E.}},
\bauthor{\bsnm{Stewart}, \binits{Z.J.}},
\bauthor{\bsnm{Weidman}, \binits{J.D.}}, \betal:
\batitle{Coulomb interaction-driven entanglement of electrons on helium}.
\bjtitle{PRX Quantum}
\bvolume{5}(\bissue{3}),
\bfpage{030324}
(\byear{2024})
\end{barticle}
\endbibitem

\bibitem[\protect\citeauthoryear{Lyon}{2006}]{lyon2006spin}
\begin{barticle}
\bauthor{\bsnm{Lyon}, \binits{S.}}:
\batitle{Spin-based quantum computing using electrons on liquid helium}.
\bjtitle{Physical Review A—Atomic, Molecular, and Optical Physics}
\bvolume{74}(\bissue{5}),
\bfpage{052338}
(\byear{2006})
\end{barticle}
\endbibitem

\bibitem[\protect\citeauthoryear{Beysengulov
  et~al.}{2022}]{beysengulov2022helium}
\begin{barticle}
\bauthor{\bsnm{Beysengulov}, \binits{N.}},
\bauthor{\bsnm{Mikolas}, \binits{C.}},
\bauthor{\bsnm{Kitzman}, \binits{J.}},
\bauthor{\bsnm{Lane}, \binits{J.}},
\bauthor{\bsnm{Edmunds}, \binits{D.}},
\bauthor{\bsnm{Rees}, \binits{D.}},
\bauthor{\bsnm{Henriksen}, \binits{E.}},
\bauthor{\bsnm{Lyon}, \binits{S.}},
\bauthor{\bsnm{Pollanen}, \binits{J.}}:
\batitle{Helium surface fluctuations investigated with superconducting coplanar
  waveguide resonator}.
\bjtitle{Journal of Low Temperature Physics}
\bvolume{208}(\bissue{5}),
\bfpage{482}--\blpage{491}
(\byear{2022})
\end{barticle}
\endbibitem

\bibitem[\protect\citeauthoryear{Zavyalov et~al.}{2005}]{zavyalov2005electron}
\begin{barticle}
\bauthor{\bsnm{Zavyalov}, \binits{V.}},
\bauthor{\bsnm{Smolyaninov}, \binits{I.}},
\bauthor{\bsnm{Zotova}, \binits{E.}},
\bauthor{\bsnm{Borodin}, \binits{A.}},
\bauthor{\bsnm{Bogomolov}, \binits{S.}}:
\batitle{Electron states above the surfaces of solid cryodielectrics for
  quantum-computing.}
\bjtitle{Journal of low temperature physics}
\bvolume{138},
\bfpage{415}--\blpage{420}
(\byear{2005})
\end{barticle}
\endbibitem

\bibitem[\protect\citeauthoryear{Zhou et~al.}{2022}]{zhou2022single}
\begin{barticle}
\bauthor{\bsnm{Zhou}, \binits{X.}},
\bauthor{\bsnm{Koolstra}, \binits{G.}},
\bauthor{\bsnm{Zhang}, \binits{X.}},
\bauthor{\bsnm{Yang}, \binits{G.}},
\bauthor{\bsnm{Han}, \binits{X.}},
\bauthor{\bsnm{Dizdar}, \binits{B.}},
\bauthor{\bsnm{Li}, \binits{X.}},
\bauthor{\bsnm{Divan}, \binits{R.}},
\bauthor{\bsnm{Guo}, \binits{W.}},
\bauthor{\bsnm{Murch}, \binits{K.W.}}, \betal:
\batitle{Single electrons on solid neon as a solid-state qubit platform}.
\bjtitle{Nature}
\bvolume{605}(\bissue{7908}),
\bfpage{46}--\blpage{50}
(\byear{2022})
\end{barticle}
\endbibitem

\bibitem[\protect\citeauthoryear{Zhou et~al.}{2024}]{zhou2024electron}
\begin{barticle}
\bauthor{\bsnm{Zhou}, \binits{X.}},
\bauthor{\bsnm{Li}, \binits{X.}},
\bauthor{\bsnm{Chen}, \binits{Q.}},
\bauthor{\bsnm{Koolstra}, \binits{G.}},
\bauthor{\bsnm{Yang}, \binits{G.}},
\bauthor{\bsnm{Dizdar}, \binits{B.}},
\bauthor{\bsnm{Huang}, \binits{Y.}},
\bauthor{\bsnm{Wang}, \binits{C.S.}},
\bauthor{\bsnm{Han}, \binits{X.}},
\bauthor{\bsnm{Zhang}, \binits{X.}}, \betal:
\batitle{Electron charge qubit with 0.1 millisecond coherence time}.
\bjtitle{Nature Physics}
\bvolume{20}(\bissue{1}),
\bfpage{116}--\blpage{122}
(\byear{2024})
\end{barticle}
\endbibitem

\bibitem[\protect\citeauthoryear{Albrecht et~al.}{1993}]{albrecht1993annealing}
\begin{barticle}
\bauthor{\bsnm{Albrecht}, \binits{U.}},
\bauthor{\bsnm{Evers}, \binits{P.}},
\bauthor{\bsnm{Leiderer}, \binits{P.}}:
\batitle{Annealing behavior of quench-condensed hydrogen and deuterium films}.
\bjtitle{Surface science}
\bvolume{283}(\bissue{1-3}),
\bfpage{419}--\blpage{422}
(\byear{1993})
\end{barticle}
\endbibitem

\bibitem[\protect\citeauthoryear{Sohaili et~al.}{2005}]{sohaili2005triple}
\begin{barticle}
\bauthor{\bsnm{Sohaili}, \binits{M.}},
\bauthor{\bsnm{Klier}, \binits{J.}},
\bauthor{\bsnm{Leiderer}, \binits{P.}}:
\batitle{Triple-point wetting of molecular hydrogen isotopes}.
\bjtitle{Journal of Physics: Condensed Matter}
\bvolume{17}(\bissue{9}),
\bfpage{415}
(\byear{2005})
\end{barticle}
\endbibitem

\bibitem[\protect\citeauthoryear{Obrien and
  Witteborn}{1984}]{obrien1984thermal}
\begin{botherref}
\oauthor{\bsnm{Obrien}, \binits{K.}},
\oauthor{\bsnm{Witteborn}, \binits{F.C.}}:
Thermal contacts between metal and glass for use at cryogenic temperatures.
NASA Technical Memorandum 85856,
NASA
(1984)
\end{botherref}
\endbibitem

\bibitem[\protect\citeauthoryear{Mack et~al.}{2007}]{Mack:07}
\begin{barticle}
\bauthor{\bsnm{Mack}, \binits{A.H.}},
\bauthor{\bsnm{Riordon}, \binits{J.}},
\bauthor{\bsnm{Dean}, \binits{C.R.}},
\bauthor{\bsnm{Talbot}, \binits{R.}},
\bauthor{\bsnm{Gervais}, \binits{G.}}:
\batitle{Local control of light polarization with low-temperature fiber
  optics}.
\bjtitle{Opt. Lett.}
\bvolume{32}(\bissue{11}),
\bfpage{1378}--\blpage{1380}
(\byear{2007})
\doiurl{10.1364/OL.32.001378}
\end{barticle}
\endbibitem

\bibitem[\protect\citeauthoryear{Sladkov et~al.}{2011}]{10.1063/1.3574217}
\begin{barticle}
\bauthor{\bsnm{Sladkov}, \binits{M.}},
\bauthor{\bsnm{Bakker}, \binits{M.P.}},
\bauthor{\bsnm{Chaubal}, \binits{A.U.}},
\bauthor{\bsnm{Reuter}, \binits{D.}},
\bauthor{\bsnm{Wieck}, \binits{A.D.}},
\bauthor{\bsnm{Wal}, \binits{C.H.}}:
\batitle{{Polarization-preserving confocal microscope for optical experiments
  in a dilution refrigerator with high magnetic field}}.
\bjtitle{Review of Scientific Instruments}
\bvolume{82}(\bissue{4}),
\bfpage{043105}
(\byear{2011})
\doiurl{10.1063/1.3574217}
{\href{https://arxiv.org/abs/https://pubs.aip.org/aip/rsi/article-pdf/doi/10.1063/1.3574217/13461222/043105\_1\_online.pdf}{{https://pubs.aip.org/aip/rsi/article-pdf/doi/10.1063/1.3574217/13461222/043105\_1\_online.pdf}}}
\end{barticle}
\endbibitem

\bibitem[\protect\citeauthoryear{Phoenix et~al.}{2020}]{10.1063/5.0012174}
\begin{barticle}
\bauthor{\bsnm{Phoenix}, \binits{J.}},
\bauthor{\bsnm{Gaudreau}, \binits{L.}},
\bauthor{\bsnm{Korkusinski}, \binits{M.}},
\bauthor{\bsnm{Zawadzki}, \binits{P.}},
\bauthor{\bsnm{Bogan}, \binits{A.}},
\bauthor{\bsnm{Studenikin}, \binits{S.}},
\bauthor{\bsnm{Williams}, \binits{R.L.}},
\bauthor{\bsnm{Sachrajda}, \binits{A.S.}}:
\batitle{{Full polarization control of fiber-delivered light in a dilution
  refrigerator}}.
\bjtitle{Review of Scientific Instruments}
\bvolume{91}(\bissue{8}),
\bfpage{083107}
(\byear{2020})
\doiurl{10.1063/5.0012174}
{\href{https://arxiv.org/abs/https://pubs.aip.org/aip/rsi/article-pdf/doi/10.1063/5.0012174/16023298/083107\_1\_online.pdf}{{https://pubs.aip.org/aip/rsi/article-pdf/doi/10.1063/5.0012174/16023298/083107\_1\_online.pdf}}}
\end{barticle}
\endbibitem

\bibitem[\protect\citeauthoryear{Grechushnikov}{1961}]{grechushnikov1961quartz}
\begin{barticle}
\bauthor{\bsnm{Grechushnikov}, \binits{B.}}:
\batitle{Quartz circular polarizers}.
\bjtitle{Optics and Spectroscopy}
\bvolume{12},
\bfpage{69}
(\byear{1961})
\end{barticle}
\endbibitem

\bibitem[\protect\citeauthoryear{MacNeille}{1946}]{MacNeille}
\begin{botherref}
\oauthor{\bsnm{MacNeille}, \binits{S.M.}}:
Beam Splitter.
2,403,731,
1946
\end{botherref}
\endbibitem

\bibitem[\protect\citeauthoryear{Hale and Day}{1988}]{Hale:88}
\begin{barticle}
\bauthor{\bsnm{Hale}, \binits{P.D.}},
\bauthor{\bsnm{Day}, \binits{G.W.}}:
\batitle{Stability of birefringent linear retarders(waveplates)}.
\bjtitle{Appl. Opt.}
\bvolume{27}(\bissue{24}),
\bfpage{5146}--\blpage{5153}
(\byear{1988})
\doiurl{10.1364/AO.27.005146}
\end{barticle}
\endbibitem

\bibitem[\protect\citeauthoryear{Pezzaniti and Chipman}{1994}]{Pezzaniti:94}
\begin{barticle}
\bauthor{\bsnm{Pezzaniti}, \binits{J.L.}},
\bauthor{\bsnm{Chipman}, \binits{R.A.}}:
\batitle{Angular dependence of polarizing beam-splitter cubes}.
\bjtitle{Appl. Opt.}
\bvolume{33}(\bissue{10}),
\bfpage{1916}--\blpage{1929}
(\byear{1994})
\doiurl{10.1364/AO.33.001916}
\end{barticle}
\endbibitem

\bibitem[\protect\citeauthoryear{Li and Dobrowolski}{2000}]{li2000high}
\begin{barticle}
\bauthor{\bsnm{Li}, \binits{L.}},
\bauthor{\bsnm{Dobrowolski}, \binits{J.}}:
\batitle{High-performance thin-film polarizing beam splitter operating at
  angles greater than the critical angle}.
\bjtitle{Applied Optics}
\bvolume{39}(\bissue{16}),
\bfpage{2754}--\blpage{2771}
(\byear{2000})
\end{barticle}
\endbibitem

\bibitem[\protect\citeauthoryear{Corruccini and
  Gniewek}{1961}]{corruccini1961thermal}
\begin{bbook}
\bauthor{\bsnm{Corruccini}, \binits{R.J.}},
\bauthor{\bsnm{Gniewek}, \binits{J.J.}}:
\bbtitle{Thermal Expansion of Technical Solids at Low Temperatures: A
  Compilation from the Literature}
vol. \bseriesno{29}.
\bpublisher{US Department of Commerce, National Bureau of Standards},
\blocation{Washington, D.C.}
(\byear{1961})
\end{bbook}
\endbibitem

\bibitem[\protect\citeauthoryear{Ekin}{2006}]{ekin2006experimental}
\begin{bbook}
\bauthor{\bsnm{Ekin}, \binits{J.}}:
\bbtitle{Experimental Techniques for Low-temperature Measurements: Cryostat
  Design, Material Properties and Superconductor Critical-current Testing}.
\bpublisher{Oxford university press},
\blocation{Great Clarendon Street, Oxford}
(\byear{2006})
\end{bbook}
\endbibitem

\bibitem[\protect\citeauthoryear{White}{1964}]{WHITE19642}
\begin{barticle}
\bauthor{\bsnm{White}, \binits{G.K.}}:
\batitle{Thermal expansion of silica at low temperatures}.
\bjtitle{Cryogenics}
\bvolume{4}(\bissue{1}),
\bfpage{2}--\blpage{7}
(\byear{1964})
\doiurl{10.1016/0011-2275(64)90029-3}
\end{barticle}
\endbibitem

\bibitem[\protect\citeauthoryear{Arp et~al.}{1962}]{arp1962thermal}
\begin{barticle}
\bauthor{\bsnm{Arp}, \binits{V.}},
\bauthor{\bsnm{Wilson}, \binits{J.}},
\bauthor{\bsnm{Winrich}, \binits{L.}},
\bauthor{\bsnm{Sikora}, \binits{P.}}:
\batitle{Thermal expansion of some engineering materials from 20 to 293 {K}}.
\bjtitle{Cryogenics}
\bvolume{2}(\bissue{4}),
\bfpage{230}--\blpage{235}
(\byear{1962})
\end{barticle}
\endbibitem

\bibitem[\protect\citeauthoryear{Toyoda and Yabe}{1983}]{Toyoda_1983}
\begin{barticle}
\bauthor{\bsnm{Toyoda}, \binits{T.}},
\bauthor{\bsnm{Yabe}, \binits{M.}}:
\batitle{The temperature dependence of the refractive indices of fused silica
  and crystal quartz}.
\bjtitle{Journal of Physics D: Applied Physics}
\bvolume{16}(\bissue{5}),
\bfpage{97}
(\byear{1983})
\doiurl{10.1088/0022-3727/16/5/002}
\end{barticle}
\endbibitem

\bibitem[\protect\citeauthoryear{Malitson}{1962}]{Malitson:62}
\begin{barticle}
\bauthor{\bsnm{Malitson}, \binits{I.H.}}:
\batitle{Refraction and dispersion of synthetic sapphire}.
\bjtitle{J. Opt. Soc. Am.}
\bvolume{52}(\bissue{12}),
\bfpage{1377}--\blpage{1379}
(\byear{1962})
\doiurl{10.1364/JOSA.52.001377}
\end{barticle}
\endbibitem

\bibitem[\protect\citeauthoryear{Born et~al.}{1999}]{Born1999}
\begin{bbook}
\bauthor{\bsnm{Born}, \binits{M.}},
\bauthor{\bsnm{Wolf}, \binits{E.}},
\bauthor{\bsnm{Bhatia}, \binits{A.B.}},
\bauthor{\bsnm{Clemmow}, \binits{P.C.}},
\bauthor{\bsnm{Gabor}, \binits{D.}},
\bauthor{\bsnm{Stokes}, \binits{A.R.}},
\bauthor{\bsnm{Taylor}, \binits{A.M.}},
\bauthor{\bsnm{Wayman}, \binits{P.A.}},
\bauthor{\bsnm{Wilcock}, \binits{W.L.}}:
\bbtitle{Principles of Optics: Electromagnetic Theory of Propagation,
  Interference and Diffraction of Light}.
\bpublisher{Cambridge University Press}, \blocation{???}
(\byear{1999}).
\doiurl{10.1017/cbo9781139644181} .
\burl{http://dx.doi.org/10.1017/CBO9781139644181}
\end{bbook}
\endbibitem

\bibitem[\protect\citeauthoryear{}{}]{casix}
\begin{botherref}
{CASIX Inc.}
\url{https://casix.com/}
\end{botherref}
\endbibitem

\bibitem[\protect\citeauthoryear{}{}]{colorPol}
\begin{botherref}
How {colorPol}\textsuperscript{\textregistered} polarizers work.
Technical report,
CODIXX.
\url{https://www.codixx.com/fileadmin/codixx_content/pdfs/PDF_sonstige/Remake_Physik_und_Know_How.pdf}
\end{botherref}
\endbibitem

\end{thebibliography}

\end{document}